\documentclass[12pt]{article}
\usepackage{graphicx,color}
\newcommand{\fr}[2]{{\hbox{$ #1 \over #2 $}}}

\begin{document}
\centerline{\hfill NSF-KITP-13-102}
\vskip 1in
\begin{center}
{\large\bf 
Superheavy Particle Origin of IceCube  PeV Neutrino Events
}
\vskip 2cm

Vernon Barger{$^{a,b}$} and Wai-Yee Keung{$^{c}$}
\vskip1cm 
$^a$Physics Department, University of Wisconsin, Madison, WI 53706 
\vskip.35cm
$^{b}$Kavli Institute for Theoretical Physics, University of California,\\ 
Santa Barbara CA 93106 
\vskip.35cm
$^{c}$Department of Physics, University of Illinois at Chicago,
Chicago, IL  60607-7059
\end{center}

\vskip0.5in
\begin{center}
{\large\bf 
ABSTRACT
} \end{center}
\vskip 1cm 
We interpret the PeV shower events observed by the IceCube
collaboration as an $s$-channel enhancement of neutrino-quark scattering
by a leptoquark that couples to the $\tau$-flavor and light quarks.
With a leptoquark mass around  $ 0.6$ TeV and a steep  $1/E^{2.3}$
neutrino flux, charged-current scattering gives cascade events at $\sim$ 1
PeV and neutral-current scattering gives cascade events at $\sim$ 0.5 PeV.
This mechanism is also consistent with the paucity of muon-track
events above 100 TeV.

\vskip3in

The IceCube (IC) experiment
has reported
intriguing results from a search
for high-energy neutrino events with a contained vertex\cite{Aartsen:2013bka,whitehorn}.
By requiring anti-coincidence with the detector edges, the
background from charged-muon initiated events can be filtered out. This allows
the observation of neutrinos from $4\pi$ directions in a fiducial volume
of 420 Megatons. The search is sensitive to all neutrino types at
energies above 50 TeV.  The charged current (CC) interactions of
muon-type neutrinos give a track associated with the muon.  The CC of
electron-type neutrinos,  the CC of
$\tau$-type neutrinos with hadronic decays of the $\tau$, and the
neutral current (NC) neutrino events give showers in the detector.

In 662 days of data, the IceCube collaboration found two spectacular
shower events with electromagnetic-equivalent energy deposits in the
detector of about 1.05 PeV and 1.15 PeV\cite{Aartsen:2013bka}.    Both
events are down-going.  A follow-up analysis\cite{whitehorn} found 26 more events with
energies between 20 TeV and 300 TeV (21 showers without tracks and 7
events with tracks indicating visible muons).  This rate  is about
twice that expected from neutrinos of charm origin from atmospheric
neutrinos\cite{Enberg}
and the energy spectrum merges well with the
atmospheric neutrino data at lower energies.  The overall signal is
inconsistent at 4.3 $\sigma$ with standard atmospheric neutrino backgrounds.
Moreover, the data suggest a potential upper cutoff at $\sim$ 2 PeV. These
observations motivate consideration of a new neutrino physics
component at ultra-high energies (UHE).

The following characteristics of the IC data focus our model
considerations:

\begin{itemize}
\item[(i)] The two PeV events are showers, which indicates
electron-neutrino CC scattering, $\nu_\tau$ CC
production of $\tau$-leptons that decay to hadrons, or NC events; 
\item[(ii)] The two PeV events have about the same energy, within
their energy uncertainties; 
\item[(iii)] The two PeV events are downward; 
\item[(iv)] No events are observed in a gap between 0.3 PeV and 1 PeV;  
we subsequently refer to this as the energy gap;
\item[(v)] Between 0.15 PeV
and 0.3 PeV, there are 2 upward showers, 2 downward showers and one
upward muon-track event.
\end{itemize}

The absence of events above 2 PeV could be the consequence of the decline in the neutrino flux from the cosmic acceleration mechanism of cosmic ray protons and iron\cite{Berezinsky}. It is possible that
above-PeV events will be observed in future 
data\cite {Kistler:2013my}.   
More exciting, from a particle physics standpoint, is that there is
indeed an approximate effective energy cut-off somewhat above 1 PeV.
It is the consequences of this, and an associated neutrino flavor
problem, that we pursue.  The flavor problem is a paucity of
muon-track events in comparison with shower events in the IceCube
events at the highest energies. 
The neutrino flux ratios at the source are converted by neutrino
oscillations to universal 1:1:1 composition\cite{LP}.
But, no muon-track events are seen above 0.3 PeV. The
track event in the 0.15 to 0.3 PeV range could be from a $\tau$ lepton
that decays to a muon.
As a cautionary remark, we note that the IceCube result is not
inconsistent with a 1:1:1 flavor composition.  The IC event selection
in the current analysis is based on deposited energy, which is more
favorable to cascade events than contained track events.  Also,
muon-neutrino events may be underrepresented because the produced muon
carries away energy.  Since the IC acceptance corrected exposure at
$\sim 1$ PeV is higher for $\nu_e$ than that for $\nu_\mu$ or
$\nu_\tau$, it is possible that the PeV events are CC interactions of
electron-neutrinos \cite{Laha:2013lka} and do not require new
physics.  Here we pursue the interesting possibility that the PeV
events are the first signals of new physics, namely a third generation
leptoquark.  

Aside from an unexpected modification to the primary neutrino flux
composition, there are several new physics possibilities that could
explain such a energy cut-off and the observed flavor asymmetry, as we
now discuss.

One such possibility is that the neutrino cross-section has a resonant
enhancement at $ \sim$ 1 PeV.  
We note that the Glashow resonance\cite{Glashow:1960zz}  at 6.3 PeV
electron-antineutrino energy is a candidate, where the
hadronic decays of the produced $W$-boson give shower events
Even so, the Glashow resonance energy is on the high side
compared to the 1 PeV of the observed shower energies.  The Glashow resonance
does not readily explain the energy gap noted above. Also, this interpretation
requires an enhanced anti-electron-neutrino flux, possibility from
decays of cosmic neutrons produced in the inelastic scatter of protons
of iron on the CMB\cite{Berezinsky}. The Glashow resonance
option has been considered
elsewhere\cite{Barger:2012mz,Bhattacharya:2012fh} and it is not the
subject of our current interest.

Another resonance enhancement candidate is neutrino scattering on
light quarks through via a leptoquark of mass $\sim 0.5$ TeV.  This is the
specific case that we shall pursue in some depth.  In particular, we shall consider a leptoquark that couples to $\tau$-lepton and  down-quark flavors.

Alternatively, the source could be a spectral line of definite
neutrino energy that could arise from the annihilations of Majorana
dark matter particles (of PeV dark matter mass) to a two-neutrino
final state (or Dirac dark matter annihilations to a neutrino and
antineutrino) or a two-body decay of a dark matter particle to a final
state with a neutrino.  The dark matter decay option has been
advocated in Ref.\cite{Feldstein:2013kka}.  

Still another scenario, that is relevant to the neutrino
flavor problem, is that the most massive neutrinos may decay to the
lighter neutrinos over cosmological distances\cite{Pakvasa:2012db}. 
This could explain the low flux of ultra UHE astrophysical
muon-neutrinos.  Then only the lightest mass eigenstate would
survive and all events would be electron-neutrino initiated showers.
The energy gap is not explained.

All of the above are conceivable new physics interpretations of the
anomalous IceCube events.  The dark matter scenarios allow a wide
range of model freedom.  The leptoquark scenario is more specifically
defined and we consider its attributes as an exemplary case, but many
of our arguments may apply more generally.  We promote the case of
a leptoquark with $\tau$-lepton and d-quark flavor.

A crux of our argument is that the Earth is almost opaque to
electron-neutrinos and muon-neutrinos.  $\tau$-neutrinos are regenerated
via $\tau$-decays, so upward neutrinos that pass through the Earth
should be of $\tau$-flavor\cite{TM}.  Since $\tau$ decays to electrons or
hadrons 82\% of the time, the $\nu_\tau$ events are dominantly
characterized as showers and only 18\% will give a muon-track.

Let us now compare this general expectation with the IceCube event
sample.  For the ensemble of events above 0.02 PeV, 
there are fewer events {upward}
than {downward},
as would be anticipated from the absorption of electron-neutrinos and
muon-neutrinos by the Earth.  Moreover, all but one of the muon-track
events are {upward or horizontal},
as expected.  However, above
0.15 PeV, there is only one muon-track event (and it is 
{upward})
compared to 6 shower events 
({2 upward and 4 downward}).
Although the statistics are low, the IceCube data suggest that 
{mainly}
$\nu_\tau$ events are being seen above 0.15 PeV.  This will be
the premise of our speculation as to their origin.

For the leptoquark (LQ) model, we assume a weak-isospin LQ doublet
that couples to third generation leptons ($\nu_\tau, \tau$) and
first and second generation quarks ($u,d$).  Thus, the main processes of
interest, because their cross-sections are resonance enhanced, are

$$ \nu_\tau  + q  \to \hbox{ LQ }  \to  \tau + q’   \ , $$
$$ \nu_\tau  + q  \to \hbox{ LQ } \to   \nu_\tau + q \ . $$
as illustrated in Fig. 1.  
\begin{figure}
\begin{center}
\includegraphics[width=4in]{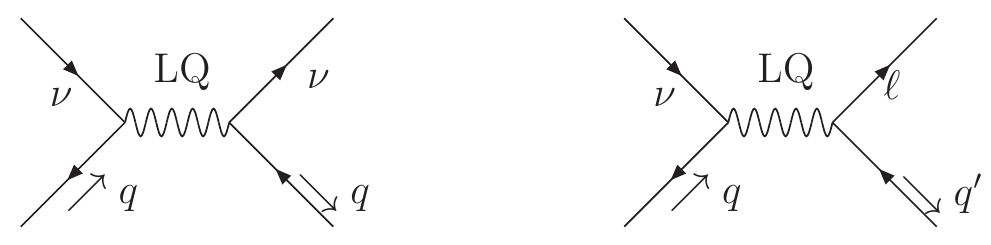}
\end{center}
\caption{Resonance processes via a leptoquark LQ 
in the UHE neutrino nucleon scattering.
Left: the neutral current events. Right: the charged current events.}
The case of interest is a $\tau$-neutrino and a $\tau$-lepton.
\end{figure}


We attribute the IC shower events at PeV energy to the CC reaction for
which the showers are associated with the $\tau$ decays to hadrons and
the hadron jet from the produced quark.  The energy of the secondary
neutrino from the $\tau$ decay is undetected, so the observed EM shower
energy is a little less than the mass of the leptoquark.  To a
zero-level approximation, the shower energy deposition determines the leptoquark mass.

When the produced $\tau$ decays to a muon, giving a track, the energy of
the event is lower than for the hadronic $\tau$-decays.  Likewise, in the
case that the $\tau$ decays to an electron the shower energy is lower
than for the $\tau$ to hadrons decay.

In the NC reaction above, the energy of the event will
be about half of the CC reaction. Thus, the shower
energy of the NC is an approximate measure of $\fr12$  the leptoquark mass.
The NC cross section is about the same as the CC cross-section.  The
gap between the PeV events and the onset of the lower energy events
should be about $\fr12$ of the leptoquark mass, which seems consistent with
what is observed.

The leptoquark can be a a scalar ($J=0$) or a vector ($J=1$). 
A general list of leptoquark models and the experimental limits are given in the review by
S. Rolli and M. Tanabashi\cite{Rolli:2008zz} in the Particle Data Book.
We show the
simple scenario of a leptoquark scalar $S$
of charge $-\fr13$, 
which couples to the first generation quarks and the third 
generation lepton in the following form,
\begin{equation}
{\cal L}_{\rm LQ}=
f_L S^\dagger  (u,d)_L \ \varepsilon\   
\left(\begin{array}{c} \nu_\tau\\ \tau\end{array}\right)_L 
+ f_R S^\dagger u_R \tau_R  + \hbox{ h.c. } \end{equation}
The Levi-Civita symbol $\varepsilon$ antisymmetrizes 
the two $SU(2)$ doublets to match the singlet $S$. 
The couplings $f_L, f_R$ are the leptoquark couplings 
to the left and right chiral quarks.  
In the narrow width approximation, 
the  leptoquark resonance contribution to
the neutrino cross-section has the 
form\cite{Anchordoqui:2006wc,Alikhanov:2013fda,Doncheski:1997it} 
\begin{equation}
  M_S^2 \ d\sigma(\nu_\tau N \stackrel{{\rm LQ}}{\longrightarrow} \nu_\tau X)
=\fr{\pi}{2} f_L^2 \ {\rm Br}( S\to \nu_\tau d) x d_N(x,\mu^2)    
\ ,\end{equation}
\begin{equation}
  M_S^2 \ d\sigma(\nu_\tau N \stackrel{{\rm LQ}}{\longrightarrow} \tau X)
=\fr{\pi}{2} f_L^2 \ {\rm Br}( S\to \tau_{L,R} u) x d_N(x,\mu^2)    
\ ,\end{equation}
where the parton fractional momentum is $x= M_S^2/s$ with $s=2m_N E_\nu$.  
The down-quark parton distribution function $d_N(x,\mu^2)$   
in the target nucleon $N$   is evaluated at the scale  $\mu^2= M_S^2$ 
in the leading order calculation. 
The inelasticity  $y = (E_\nu-E')/E_\nu$ distribution 
is flat in the scalar leptoquark scenario; here $E'$ denotes the outgoing energy of $\nu_\tau$ or $\tau$. 
The threshold energy of LQ production is $E_\nu = M_S^2/2m_N$.

The branching fractions are
\begin{eqnarray}
{\rm Br}( S\to \nu_\tau d) &=&f_L^2/(2f_L^2+f_R^2)  \ ,\nonumber  \\
{\rm Br}( S\to \tau_L u) &=&f_L^2/(2f_L^2+f_R^2)  \ ,\nonumber \\
{\rm Br}( S\to \tau_R u) &=&f_R^2/(2f_L^2+f_R^2)  \ . \end{eqnarray}
They multiply the leptoquark production cross-section (see Fig.~2),
\begin{equation} 
\sigma_{\rm LQ}(\nu N)=\frac{\pi f_L^2}{2 M_S^2} x d_N (x, \mu^2) \ , 
\end{equation}
to produce the corresponding rates for each channel.  For a vector leptoquark
the cross section is a factor of 2 larger than that of the scalar leptoquark.
\begin{figure}
\begin{center}
\includegraphics[width=5in,height=4in]{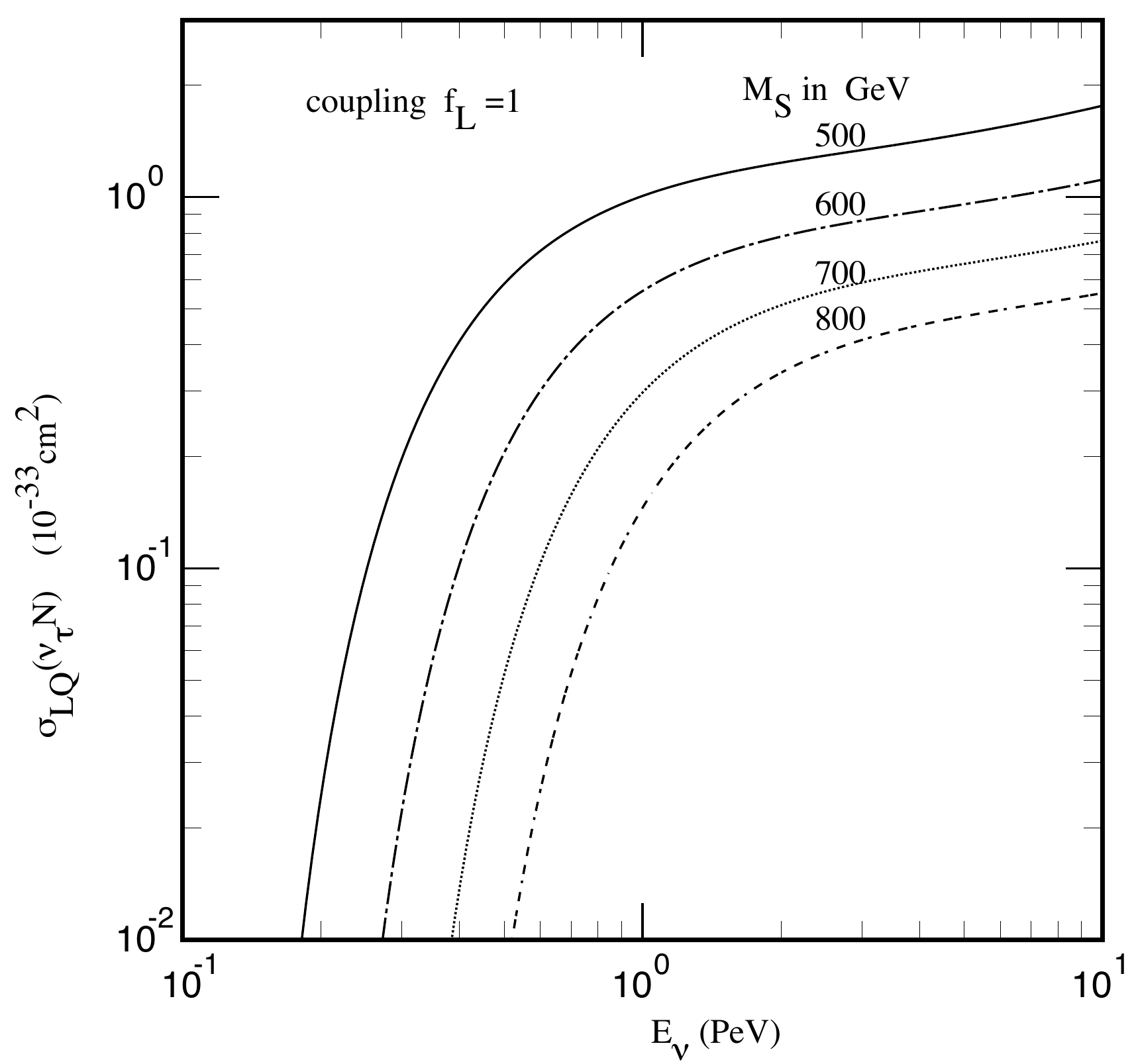}
\end{center}
\caption{LQ production cross-section in $\nu_\tau N$ scattering.}
\end{figure}

The LQ width is given by
\begin{equation}  \Gamma_{\rm LQ}=\fr1{16\pi}M_S (2f_L^2+f_R^2)  \ , \end{equation}
which is a small fraction of its mass even for a unit coupling $f$, 
so the narrow width approximation is justifiable.  
The partial widths of  $S\to \nu_\tau +d $ and 
$S \to\tau_L +u $ are equal. However, $f_R$ gives rise to the channel
$S \to\tau_R +u $.

Figure 1 shows the cross section of scalar $\tau-q$ leptoquark production in neutrino  scattering on an isospin averaged nucleon
target N, 
taking $f_L$ =1. The CTEQ6.10 parton distributions at NLO are used in this calculation\cite{CTEQ}.

As a benchmark of the  neutrino flux, we adopt for illustration the
A-W\cite{Anchordoqui:2013qsi} form with a steep  power index 
$\Gamma=2.3$ based on the optimal fitting with 
the minimal deviation from IceCube UHE neutrino data. 
\begin{equation} 
 \Phi^{\rm A-W}_\nu =
\Phi_0 \left({E_\nu\over 1\  {\rm GeV}}\right)^{-\Gamma} \ , \ 
\Phi_0=6.62 \times 10^{-7}\ {\rm /GeV/cm^2/s/sr} \ . 
\end{equation}
for each neutrino-type.  We estimate the expected event number at IceCube by
$$ {\cal N}=  n t \Omega \int dE_{\nu} \sigma_{\rm LQ} 
\cdot [ {\rm Br} ] \cdot
\Phi^{\rm A-W}_{\nu}(E_\nu) \ , $$
where we take the time of exposure $t=662$ days, 
the effective target nucleons number in IceCube $n=6\times 10^{38}$,
and the solid angle of the full $4\pi$ coverage ($\Omega=4\pi$).
\begin{figure}
\begin{center}
\includegraphics[width=6in]{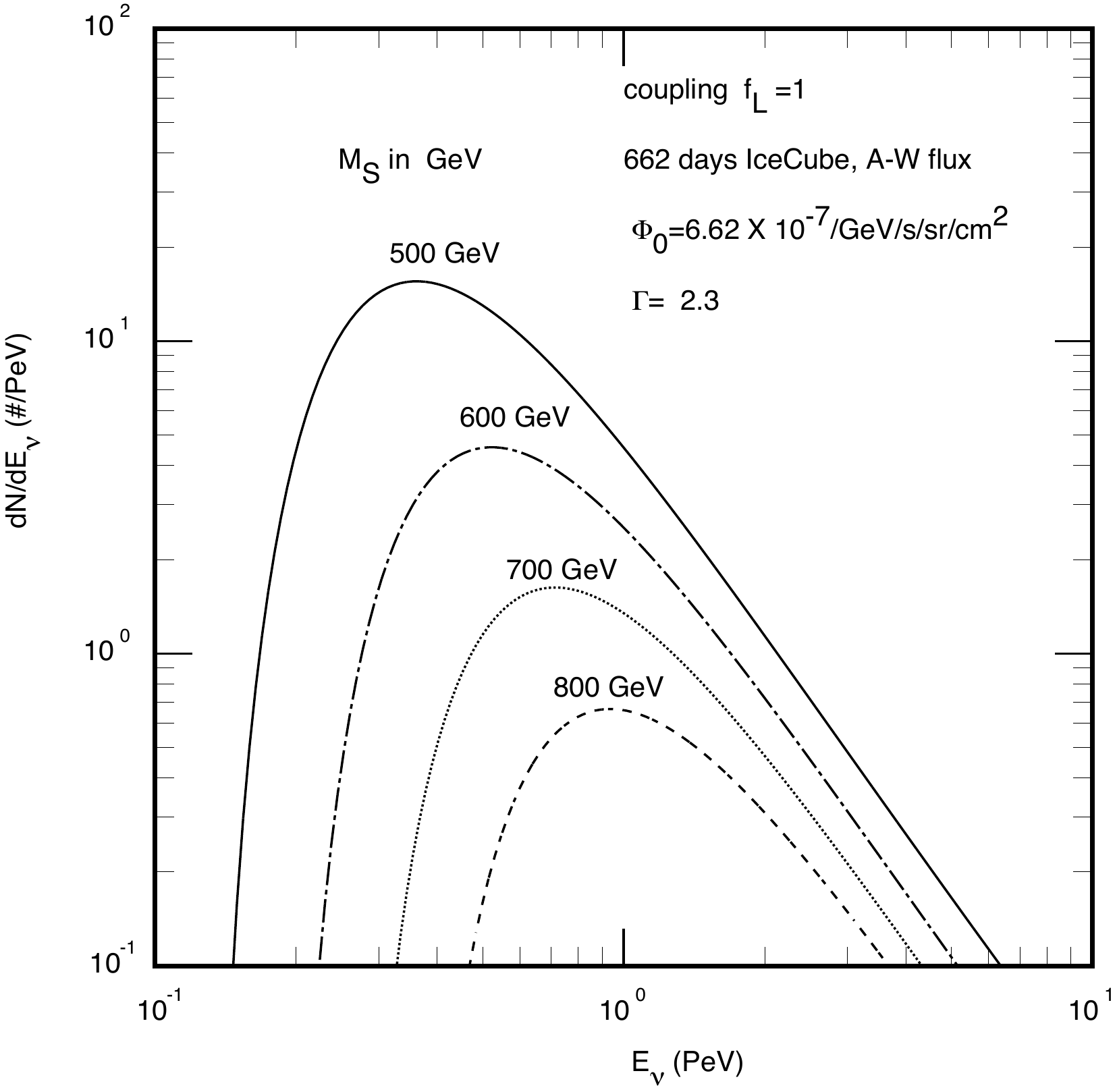}
\end{center}
\caption{Event rate distribution $d{\cal N}/dE_\nu$, from the LQ cross section
convoluted with the {A-W} flux $\Phi^{\rm A-W}_\nu$.}
\end{figure}
The event distribution $d{\cal N}/dE_\nu$ is given in Fig.~3.  
Below we tabulate  the LQ event rates in three $E_\nu$  bins,
for two LQ masses $M_S$.
\begin{equation}
   \begin{array}{||c|c|c|c||} \hline
    M_S ({\rm GeV})  &   
    < {\rm\ 1 PeV}  &    {\rm 1 -2 PeV} &   > {\ \rm 2 PeV} \\ \hline
    500  &     8.2  &    2.3  &    1.8  \\  
    600  &     2.6  &    1.4  &    1.1  \\  \hline  
   \end{array} \ . \end{equation}
At a neutrino energy of $\sim$ 1 PeV a few events are predicted for LQ mass
$\sim$ 0.5 TeV, in rough accord with the two shower events observed by
IceCube.

Because the cosmic neutrino flux is unknown, 
we also show the ratio of the CC $\tau$-cross section with the 
LQ resonance to the Standard Model CC $\tau$-cross section in Fig. 4.  
This figure demonstrates the enhancement of $\tau$ events by the LQ scenario 
without assumptions about the flux.  
\begin{figure}
\begin{center}
\includegraphics[width=6in,height=4in]{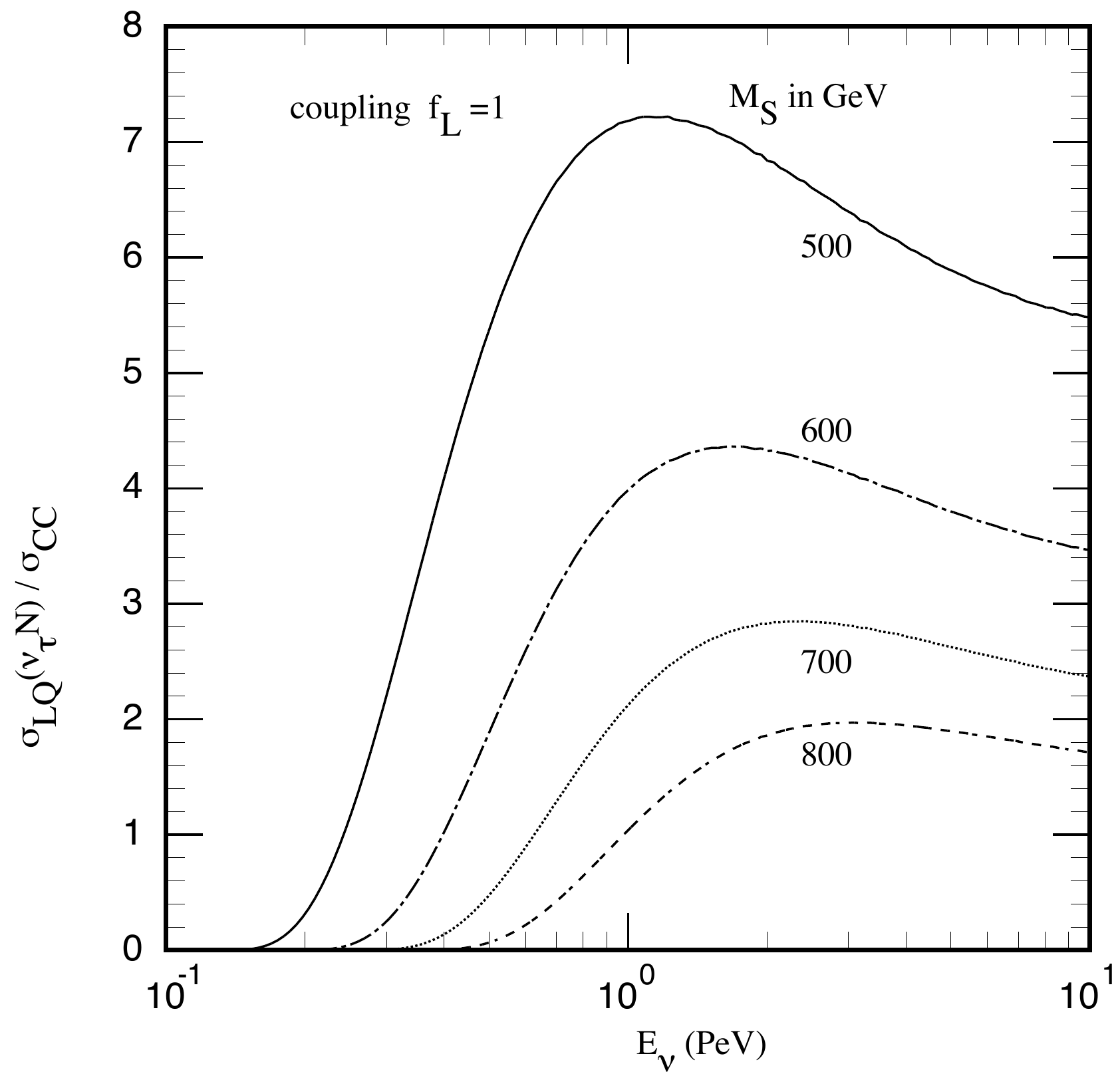}
\end{center}
\caption{Cross section ratio of the charged current process of
$\nu_\tau N$ scattering via a leptoquark resonance to the  
corresponding Standard Model CC process.}
\end{figure}

The LQ model illustrations above are for the coupling choice $f_L =1$;
the LQ cross section scales with the value of $f_L^2$.  According to
Fig.~3, with $f_L =1$, a LQ mass of 600 GeV would lead to 1 to 2 cascade
events with reconstructed neutrino energies of order 1 PeV; the number
of LQ events above 1 PeV falls rapidly with energy due to the
convolution with the assumed $E^{-2.3}$ flux.

It is appropriate to ask whether the LHC can probe the LQ coupling and
LQ mass that could account for the IceCube PeV cascade events.  
The LQ production cross sections at LHC are calculated in 
Ref.\cite{CiezaMontalvo:1998sk, Belyaev:2005ew}.
Based on its pair production, the CMS/LHC search 
at 7 TeV\cite{Chatrchyan:2012sv} 
for a scalar $\tau$-type LQ puts a constraint 
$M_S \stackrel{>}{\sim} 525$ GeV.  Single LQ
production at the LHC can occur through the subprocesses
$$ g u \to \bar\tau S \ ,\quad gd \to \bar\nu_\tau S \ . $$
The down-type LQ, $S$, subsequently decays into $ \tau u$ or $\nu_\tau
d$.  The resulting final states are $\bar\tau\tau u$, or
$\bar\tau\nu_\tau d$, or $\bar\nu_\tau\nu_\tau d$, etc. Rather
distinctively, these subprocesses give events of a $\bar\tau\tau$ pair
plus a jet or a monojet and missing energy with or without a $\tau$.
A dedicated search at LHC14 for the single LQ production signals is of
great interest to confirm or deny the proposed LQ explanation  
of the PeV IceCube cascade events.

In summary, we have interpreted the UHE neutrino events
observed in the IceCube experiment in the framework of a resonance
enhancement by a leptoquark that couples to the $\tau$-lepton and to
the down-type quark.  The characteristic features of the events that
are reproduced by the model are (i) a cross-section enhancement above
atmospheric neutrino expectations, (ii) dominance of shower events
over track events above 100 TeV, interpreted as dominance of
$\nu_\tau$ processes, (iii) an energy gap between 0.3 PeV and 1 PeV
where no events were recorded, attributed to NC showers at lower
energies and $\nu_\tau$ CC showers at PeV energies.  A leptoquark
mass in the vicinity of 0.6 TeV is inferred.
The next release of data, accumulated in IceCube since 2012, should
further test the leptoquark explanation of the PeV events.

\section*{Acknowledgements}

VB thanks KITP-UCSB for hospitality during the course of this work and
F. Halzen, A. Karle, J. Learned, D. Marfatia, S. Pakvasa and 
{W.P. Pan}
for email communications.
This research was supported in part by the National Science Foundation
under Grant No. NSF PHY11-25915 and in part by the U.S. Department of
Energy under grants No. DE-FG02-95ER40896 and DE-FG02-12ER41811.

\end{document}